\documentclass[english]{article}
\usepackage[T1]{fontenc}
\usepackage[latin9]{inputenc}
\usepackage{amsmath}
\usepackage{amssymb}

\makeatletter
\usepackage{amssymb}
\usepackage{latexsym}
\usepackage{epsfig}
\usepackage{float}
\usepackage{dsfont}

\newcommand{\be}{\begin{equation}}
\newcommand{\ee}{\end{equation}}

\allowdisplaybreaks

\makeatother

\usepackage{babel}
\begin{document}
{}~ \hfill\vbox{\hbox{CTP-SCU/2017004}}\break
\vskip 3.0cm
\centerline{\Large \bf Correspondence between genus expansion and $\alpha^{\prime}$ expansion in string theory}
\vspace*{10.0ex}
\centerline{\large Peng Wang$^a$, Houwen Wu$^{a,b}$ and Haitang Yang$^a$}
\vspace*{7.0ex}
\vspace*{4.0ex}
\centerline{\large \it $^a$Center for theoretical physics}
\centerline{\large \it Sichuan University}
\centerline{\large \it Chengdu, 610064, China} \vspace*{1.0ex}
\vspace*{4.0ex}
\centerline{\large \it $^b$Center for the Fundamental Laws of Nature}
\centerline{\large \it Harvard University}
\centerline{\large \it Cambridge, MA 02138 USA} \vspace*{1.0ex}

\centerline{pengw@scu.edu.cn, wu\_houwen@g.harvard.edu, hyanga@scu.edu.cn}

\vspace*{10.0ex}
\centerline{\bf Abstract} \bigskip \smallskip
In this paper, we demonstrate that locally, the $\alpha^{\prime}$ expansion of a  string propagating in AdS can be summed into a closed expression, where the $\alpha'$ dependence is manifested. The T-dual of this sum  exactly matches the expression controlling all genus expansion in the Goparkumar-Vafa formula, which in turn also matches the loop expansion of the Chern-Simons gauge theory. We therefore find an exact correspondence between the $\alpha^{\prime}$ expansion for a string moving in AdS and the genus expansion of a string propagating in four dimensional flat spacetime. We are then able to give a closed form of the $\alpha'$ expansion for all values of $\sqrt{\alpha'}/R_{AdS}$. Moreover, the correspondence makes it possible to conjecture the exact $g_s$ dependence  of the strongly coupled  theories. 
%We also show that in AdS, the lengths of the T-dual strings satisfy $\ell_s \tilde\ell_s =R_{AdS}^2$. 

\vfill 
\eject
\baselineskip=16pt
\vspace*{10.0ex}
%\tableofcontents

\section{Motivations }

The celebrated AdS/CFT correspondence conjectures an equivalence of
the gravitational theory in the bulk of AdS and the gauge theory on
the AdS conformal boundary \cite{Maldacena:1997re}. The parameters
in these two theories are related with $\lambda=g_{{\rm YM}}^{2}N=4\pi g_{s}N=\left(\frac{R_{AdS}}{\sqrt{\alpha^{\prime}}}\right)^{4}$,
where $\lambda$ is the 't Hooft coupling, $g_{{\rm YM}}$ is the
Yang-Mills coupling, $g_{s}$ is the string coupling and $SU(N)$
is the gauge group. $\frac{R_{AdS}}{\sqrt{\alpha^{\prime}}}$ comes
from string theory where $R_{AdS}$ is the radius of the AdS background
and $\sqrt{\alpha^{\prime}}\sim\ell_{s}$ is the string length. As
one can easily see from the parameter relations, large $\lambda$
maps to small $\sqrt{\alpha'}/R_{AdS}$ and vice versa. 
%This strong-weak
%correspondence is remarkable, though makes the verification not easy.

It is well known that string theory has a double perturbation expansion.
One is the genus expansion, controlled by the string coupling $g_{s}=\exp\left(\phi\right)$,
where $\phi$ is the dilaton. This expansion is sometimes interpreted
as classical string interactions, counting worldsheets with different
topologies characterized by the genus. In the 't Hooft's argument,
this genus expansion is equivalent to the loop expansion of a Yang-Mills
gauge theory in terms of the 't Hooft coupling $\lambda=g_{{\rm YM}}^{2}N=4\pi g_{s}N$.
The second one is the $\alpha^{\prime}$ correction, coming from the
non-linear Sigma model when the string propagates in a curved background
with a characteristic radius of curvature $R_{c}$. The $\alpha'$
expansion is controlled by the parameter $\sqrt{\alpha'}/R_{c}$ and
sometimes referred as a quantum expansion.

Since the controlling parameters of the $\alpha'$ expansion and genus
expansion are related by the AdS/CFT correspondence, it is appealing
to ask if there exists some relation between these two expansions.
It is obvious that the possibly existing relation cannot arise in
a simple manner since we have a strong-weak correspondence $g_{s}^{1/4}\sim\frac{1}{\sqrt{\alpha'}/R_{AdS}}$.
In order to have such a correspondence, there ought to be a mechanism
somehow transfers a large $\sqrt{\alpha'}/R_{AdS}$ theory to a small
$\sqrt{\alpha'}/R_{AdS}$ theory, or the same pattern for the $g_{s}$
side (S-duality may play the role). 

The purpose of this paper is to show there does exist a local exact
correspondence between the $\alpha'$ expansion of a string moving
in AdS and genus expansion of a string propagating in flat spacetime.
We will show that in AdS, T-duality exhibits in a different manner
from the ordinary one. As we know, compactification is a prerequisite
for the ordinary T-duality and the physics are identical for compactification
radii $R$ and $\alpha'/R$. However, in AdS, using the Buscher's
rule \cite{Buscher:1988}, we are going to demonstrate that the lengths
of T-dual strings are related by $\ell\tilde{\ell}=R_{AdS}^{2}$ .
The difference here from the ordinary T-duality is that compactification
is not necessary and two dual strings have different string lengths
$\ell$ and $\tilde{\ell}$ respectively. It seems that this is possible
only for AdS geometry since it is maximally symmetric, conformally flat and the dual geometry
is still an AdS with the same radius. Therefore, when performing $\alpha'$
expansion, the dual strings are expanded in dual regimes, namely small/large
respectively. It is conceivable that in order to justify this T-duality,
there ought to be a correspondence between the $\alpha'$ expansion
and genus expansion. Or conversely speaking, the existence of the
correspondence must predict a duality between large $\sqrt{\alpha'}/R_{AdS}$
theory and a small $\sqrt{\alpha'}/R_{AdS}$ one.

Remarkably, we find that if we take Riemann normal coordinate to locally
expand a string moving in AdS, the $\alpha'$ expansion can be added up to a closed expression. Furthermore, the closed form of the T-dual
string is the same as the expression controlling all genus expansion
in the Goparkumar-Vafa formula (GV). With this correspondence, it
is possible to make predictions on $g_{s}$ dependence for strongly
coupled GV, i.e the strongly coupled gauge theory.

Since we are going to compare the $\alpha'$ expansion to the genus
expansion of the GV formula, it is necessary to recall some results
of the GV formula \cite{Gopakumar:1998ii,Gopakumar:1998jq,Dedushenko:2014nya}.
As we know, $F-$terms in a supersymmetric field theory or string
theory play special roles since nonperturbative results can usually
be extracted from $F-$terms only. There exist such $F-$terms when
compactifying Type IIA superstring on $\mathbb{R}^{4}\times Y$ where
$Y$ is a Calabi-Yau three-fold

\begin{equation}
\mathcal{I}=\sum_{g\ge0}I_{g}=-i\int_{\mathbb{R}^{4}}d^{4}xd^{4}\theta\sum_{g\ge0}\mathcal{F}_{g}(\mathcal{X}^{\Lambda})(\mathcal{W}_{AB}\mathcal{W}^{AB})^{g},\label{eq:F-terms}
\end{equation}
where $\theta$ is the superspace chiral coordinate, $g$ is the worldsheet
genus, $\mathcal{X}^{\Lambda}(x,\theta)=X^{\Lambda}+\theta\Psi^{\Lambda}+\theta^{2}F^{\Lambda}+\cdots$
are the $\mathcal{N}=2$ chiral superfields associated to vector multiplets,
$\mathcal{W}_{AB}$ with spinor indices $A,B$ is a superfield with
the anti-selfdual graviphoton $W^{-}$as its bottom component. A crucial
property of this interaction is that $I_{g}$ only exists in genus
$g$ order precisely and in the string frame proportional to $g_{s}^{2g-2}$.
Gopakumar and Vafa proposed that this interaction $\mathcal{I}$ should
be interpreted as the four dimensional superspace effective action
in a supersymmetric background with a constant anti-seldual graviphoton.
Moreover, instead of calculating the holomorphic function $\mathcal{F}_{g}$
and then $\mathcal{I}_{g}$ order by order, as what did in topological
string theory, we should directly calculate $\mathcal{I}$ as a whole.
This idea is achieved by lifting a Type IIA superstring compactified
on $\Sigma\subset\mathbb{R}^{4}\times Y$ to an M2-brane wrapped on
$\Sigma\times S^{1}\subset\mathbb{R}^{4}\times S^{1}\times Y$. As
follows, they give a remarkable (GV) formula

\begin{equation}
\mathcal{I}=-\int\frac{d^{4}xd^{4}\theta}{\left(2\pi\right)^{4}}\sqrt{g^{E}}e^{2\pi i\sum_{I}q_{I}\mathcal{Z}^{I}}\left[\frac{\frac{1}{4}g_{s}^{2}\mathcal{W}^{2}}{\sinh^{2}\left(\frac{1}{2}g_{s}\mathcal{W}\right)}\right]\frac{\pi}{16}^{2}\frac{1}{g_{s}^{2}},\label{eq:GV}
\end{equation}

\noindent where $g^{E}$ is the Einstein metric. The compactification
factor $e^{2\pi i\sum_{I}q_{I}\mathcal{Z}^{I}}$ determined by the
mass and constant gauge field is irrelevant to our discussion. Without
loss of generality, we take the degree $k=1$. We used the conventions
in \cite{Dedushenko:2014nya}, and replaced the $\mathcal{X}_{0}$
(imaginary) there by the string coupling $g_{s}$. As explicitly showed
in \cite{Dedushenko:2014nya}, this action includes the sum of contributions
from all string worldsheets of genus $g\geq0$, completely controlled
by the expansion:

\begin{equation}
\frac{\frac{1}{4}g_{s}^{2}\mathcal{W}^{2}}{\sinh^{2}\left(\frac{1}{2}g_{s}\mathcal{W}\right)}=1-\frac{g_{s}^{2}}{12}\mathcal{W}^{2}+\frac{g_{s}^{4}}{240}\mathcal{W}^{4}+\mathcal{O}(\mathcal{W}^{6}).\label{eq:Genus expansion}
\end{equation}

\noindent The contribution from world-sheets of genus $g$ is identified
as that proportional to $\mathcal{W}^{2g}$, which can be understood
from (\ref{eq:F-terms}). Therefore, if there exists an exact relation
between the $\alpha'$ expansion and genus expansion, this term (\ref{eq:Genus expansion})
must also serve as a closed sum of the $\alpha'$ expansion and completely
control the expansion. Indeed as we are going to show, the $\alpha'$
expansion in AdS does give

\begin{equation}
g^{ij}\left(\tilde{X}\right)=\left[\frac{\frac{1}{4}\left(\frac{\tilde{\lambda}}{4\pi}\right)^{2}\mathcal{\tilde{\mathcal{R}}}^{2}}{\sinh^{2}\left(\frac{1}{2}\frac{\tilde{\lambda}}{4\pi}\tilde{\mathcal{R}}\right)}\right]^{i}\,_{a}\,\eta^{aj},
\end{equation}
with definitions of the quantities in (\ref{eq:weak dual set}).  Moreover, it turns out that our result makes it possible to make conjectures on
the expression of the strong $g_{s}>1$ theory.

\section{T-duality in AdS background}

The ordinary T-duality demands background compactifications and says
that the physics are identical for compactified radii $R$ and $\alpha'/R$.
But in AdS geometry, we are going to show that T-duality exhibits
a different manner. We start with the usual AdS metric in Poincare
coordinates

\begin{equation}
ds^{2}=\frac{R_{AdS}^{2}}{Z^{2}}\left(dZ^{2}+\eta^{ij}dX_{i}dX_{j}\right),
\end{equation}

\noindent where the constant $R_{AdS}$ is the radius. For convenience
and clarity, we use double coordinates $(X^{i},\tilde{X}_{j})$ to
represent the T-dual fields. We take the convention that the worldsheet
coordinate $(\tau,\sigma)$ are dimensionless. As introduced in \cite{Alday:2007hr},
the T-duality transformations are

\begin{eqnarray}
g_{ij}\partial_{\sigma}X^{j} & = & \partial_{\tau}\tilde{X}_{i},\nonumber \\
g^{ij}\partial_{\sigma}\tilde{X}_{j} & = & \partial_{\tau}X^{i},\label{eq:1st EOM}
\end{eqnarray}

\noindent combined with 

\begin{equation}
\tilde{Z}\equiv\frac{R_{AdS}^{2}}{Z}.
\end{equation}

\noindent Then the dual geometry of AdS is still an AdS with the same
radius

\begin{equation}
ds^{2}=\frac{R_{AdS}^{2}}{\tilde{Z}^{2}}\left(d\tilde{Z}^{2}+\eta_{ij}d\tilde{X}^{i}d\tilde{X}^{j}\right),
\end{equation}

\noindent where the metric are defined by $g_{\mu\nu}=diag\{\frac{R_{AdS}^{2}}{Z^{2}},\frac{R_{AdS}^{2}}{Z^{2}}\eta_{ij}\}$
and the dual metric $g^{\mu\nu}=diag\{\frac{R_{AdS}^{2}}{\tilde{Z}^{2}},\frac{R_{AdS}^{2}}{\tilde{Z}^{2}}\eta^{ij}\}$.
The definitions are completely consistent as one can easily check
by using $\left(\frac{R_{AdS}^{2}}{Z^{2}}\right)^{-1}=\frac{R_{AdS}^{2}}{\tilde{Z}^{2}}$.
Let us denote the characteristic length of $X$ as $\ell_{s}\equiv\sqrt{\alpha'}$
and write $X^{i}=\ell_{s}\mathbb{X}^{i}$, where $\mathbb{X}^{i}$
is dimensionless. Similarly, we set $Z=\ell_{s}\mathbb{Z}$. Therefore,
from the T-duality (\ref{eq:1st EOM}), we can get

\begin{equation}
\tilde{X}_{j}=\int d\sigma\frac{R_{AdS}^{2}}{Z^{2}}\partial_{\tau}X^{i}=\frac{R_{AdS}^{2}}{\ell_{s}}\int d\sigma\,\mathbb{Z}^{-2}\partial_{\tau}\mathbb{X}.
\end{equation}

\noindent Since $\sigma$, $\tau$, $\mathbb{Z}$ and $\mathbb{X}$
are all dimensionless, it is natural to identify the characteristic
length of $\tilde{X}_{j}$ as $\tilde{\ell}_{s}=\frac{R_{AdS}^{2}}{\ell_{s}}$.
This identification is consistent for $\tilde{X}_{0}$ and $\tilde{Z}$
too. Therefore, the T-duality in AdS geometry takes the form:

\begin{eqnarray}
\left(Z,X^{0},X^{i}\right): &  & \ell_{s},\nonumber \\
\left(\tilde{Z},\tilde{X}_{0},\tilde{X}_{i}\right): &  & \tilde{\ell}_{s},
\end{eqnarray}

\noindent under

\begin{equation}
\ell\,\tilde{\ell}_{s}=R_{AdS}^{2}.\label{eq:T-dual relation}
\end{equation}
We can see that this T-duality relation is quite different from the
ordinary one. We do not perform any compactifications and there is
only one radius but with two string lengths. An interesting feature
is that as we know, AdS geometry is non-compact. There might be some
mathematical implications from this T-duality. Moreover, the dual
relation (\ref{eq:T-dual relation}) indicates a strong-weak correspondence,
which is very useful when we $\alpha'$ expand the string theory.

\section{$\alpha^{\prime}$ expansion of strings propagating in AdS}

In conformal gauge, the action describing a string moving in a curved
background, namely the non-linear Sigma model, is

\begin{equation}
S=-\frac{1}{4\pi\alpha'}\int_{\Sigma}g_{ij}(X)\partial_{\alpha}X^{i}\partial^{\alpha}X^{j}
\end{equation}

\subsection{Case $\ell_{s}/R_{AdS}=\sqrt{\alpha'}/R_{AdS}<1$}

In this regime, we can expand $X$ but not the dual $\tilde{X}$.
Consider expanding the field at some point $\bar{x}$,

\begin{equation}
X^{i}\left(\tau,\sigma\right)=\bar{x}^{i}+\ell_{s}\mathbb{Y}^{i}\left(\tau,\sigma\right),
\end{equation}

\noindent where $\mathbb{Y}^{i}$'s are dimensionless fluctuations.
Note the $\alpha'$ in front of the integral in the action is canceled
under this expansion. Locally around any point, we can always pick
Riemann normal coordinates such that the metric expansion is greatly
simplified

\begin{eqnarray}
g_{ij}\left(X\right) & = & \eta_{ij}+\frac{\ell_{s}^{2}}{3}R_{iklj}\mathbb{Y}^{k}\mathbb{Y}^{l}+\frac{\ell_{s}^{3}}{6}D_{k}R_{ilmj}\mathbb{Y}^{k}\mathbb{Y}^{l}\mathbb{Y}^{m}\nonumber \\
 &  & +\frac{\ell_{s}^{4}}{20}\left(D_{k}D_{l}R_{imnj}+\frac{8}{9}R_{iklp}R_{\;mnj}^{p}\right)\mathbb{Y}^{k}\mathbb{Y}^{l}\mathbb{Y}^{m}\mathbb{Y}^{n}+\ldots.
\end{eqnarray}

\noindent We now set the background as a $D$-dimensional AdS spacetime,
which is a maximally symmetric space with $D_{m}R_{ikjl}=0$ and $R_{ikjl}=-\frac{1}{R_{AdS}^{2}}\left(g_{ij}g_{kl}-g_{il}g_{kj}\right)$.
It is remarkable that, with some careful calculation, the expansion
can be summed  into a closed form,

\begin{eqnarray}
g_{ij}(X) & = & \eta_{ij}+\frac{1}{2}\sum_{n=1}^{\infty}\frac{2^{2n+2}}{(2n+2)!}\,\left(\frac{\lambda}{8\pi}\right)^{2n} \eta_{im}\left(\mathcal{R}^{2}\right)^{m}\,_{a_{1}}\left(\mathcal{R}^{2}\right)^{a_{1}}\,_{a_{2}}\cdots\left(\mathcal{R}^{2}\right)^{a_{n-1}}\,_{j}\nonumber \\
 & = & \left[\frac{\sinh^{2}\left(\frac{1}{2}\frac{\lambda}{4\pi}\mathcal{R}\right)}{\frac{1}{4}\left(\frac{\lambda}{4\pi}\right)^{2}\mathcal{R}^{2}}\right]^{a}\,_{i}\,\eta_{aj},\label{eq:weak X}
\end{eqnarray}

\noindent where we defined

\begin{equation}
\lambda\equiv\frac{\ell_{s}^{4}}{R_{AdS}^{4}},\qquad\left(\mathcal{R}^{2}\right)^{a}\,_{b}\equiv64\pi^{2}\,\,\left(\frac{R_{AdS}}{\ell_{s}}\right)^{6}R_{AdS}^{2}R^{a}\,_{ijb}\mathbb{Y}^{i}\mathbb{Y}^{j}=64\pi^{2}\,\left(\frac{R_{AdS}}{\ell_{s}}\right)^{6}\left(\delta_{b}^{a}\mathbb{Y}^{2}-\mathbb{Y}^{a}\mathbb{Y}_{b}\right).\label{eq:para define}
\end{equation}

\noindent Noted that this $\alpha'$ expansion for $X$ is valid only
for $\ell_{s}/R_{AdS}<1$. The indices are raised and lowered by $\eta_{ab}$.
Clearly it is not legal to expand the dual field $\tilde{X}$, since
the coupling $\sqrt{\tilde{\alpha}'}/R_{AdS}=\tilde{\ell}/R_{AdS}=R_{AdS}/\ell_{s}$>1.
However, the T-duality of the metric $g^{ij}(\tilde{X})=\left(g_{ij}(X)\right)^{-1}$enables
us to get a closed expression for the strongly coupled dual theory,
though in terms of $\lambda=\ell_{s}^{4}/R_{AdS}^{4}$ and variables
$\mathbb{Y}_{i}$. 

\begin{equation}
g^{ij}(\tilde{X})=\left(g_{ij}(X)\right)^{-1}=\left[\frac{\frac{1}{4}\left(\frac{\lambda}{4\pi}\right)^{2}\mathcal{R}^{2}}{\sinh^{2}\left(\frac{1}{2}\frac{\lambda}{4\pi}\mathcal{R}\right)}\right]^{i}\,_{a}\,\eta^{aj}.\label{eq:strong dual}
\end{equation}
Apparently, it looks similar to the genus expansion (\ref{eq:Genus expansion}).
The only problem is that it is not expressed by the native dual coupling
$\tilde{\lambda}=\tilde{\ell}_{s}^{4}/R_{AdS}^{4}$ and variables
$\tilde{Y}_{i}$. Actually, we can substitute $\lambda=1/\tilde{\lambda}$.
To replace $\mathbb{Y}$ by $\mathbb{\tilde{Y}}$, we can make use
of the T-duality transformation (\ref{eq:1st EOM}). We can expect
after these substitutions, the form will change and may not be exactly
the same as the genus expansion (\ref{eq:Genus expansion}). But we
do get a closed expression for a strongly coupled theory. To have
a clear comparison of the two expansions, let us consider another
regime.

\subsection{Case $\ell_{s}/R_{AdS}=\sqrt{\alpha'}/R_{AdS}>1$}

\noindent This is the regime $\tilde{\ell}/R_{AdS}<1$, so we can
expand the dual string $\tilde{X}$ by the dual length $\tilde{\ell}_{s}$

\begin{equation}
\tilde{X}_{i}\left(\tau,\sigma\right)=\tilde{\bar{x}}_{i}+\tilde{\ell}_{s}\tilde{\mathbb{Y}}_{i}\left(\tau,\sigma\right),
\end{equation}

\noindent It is worthwhile to emphasis that the native dual metric
to expand is $g^{ij}(\tilde{X)}$. Follow the same pattern, the metric
of the T-dual theory can also be put into a closed form

\begin{equation}
g^{ij}\left(\tilde{X}\right)=\left[\frac{\frac{1}{4}\left(\frac{\tilde{\lambda}}{4\pi}\right)^{2}\mathcal{\tilde{\mathcal{R}}}^{2}}{\sinh^{2}\left(\frac{1}{2}\frac{\tilde{\lambda}}{4\pi}\tilde{\mathcal{R}}\right)}\right]^{i}\,_{a}\,\eta^{aj}\label{eq:weak dual}
\end{equation}

\noindent where we set

\begin{equation}
\tilde{\lambda}\equiv\frac{\tilde{\ell}_{s}^{4}}{R_{AdS}^{4}}=\lambda^{-1},\qquad\left(\mathcal{\tilde{\mathcal{R}}}^{2}\right)^{a}\,_{b}\equiv64\pi^{2}\,\,\left(\frac{R_{AdS}}{\tilde{\ell}_{s}}\right)^{6}R_{AdS}^{2}R^{a}\,_{ijb}\mathbb{\tilde{\mathbb{Y}}}^{i}\tilde{\mathbb{Y}}^{j}=64\pi^{2}\,\left(\frac{R_{AdS}}{\tilde{\ell}_{s}}\right)^{6}\left(\delta_{b}^{a}\mathbb{\tilde{\mathbb{Y}}}^{2}-\mathbb{\tilde{\mathbb{Y}}}^{a}\tilde{\mathbb{Y}}{}_{b}\right).\label{eq:weak dual set}
\end{equation}

\noindent It is really nice to see that this $\tilde{\alpha}'$ expansion
matches the genus expansion (\ref{eq:Genus expansion}) perfectly
under the identifications

\begin{equation}
4\pi g_{s}=\tilde{\lambda}=\lambda^{-1}=\left(\frac{R_{AdS}}{\sqrt{\alpha'}}\right)^{4},\qquad\mathcal{W}=\tilde{\mathcal{R}},\label{eq:para iden}
\end{equation}
which is in perfect agreement with AdS/CFT! Again, the strongly coupled
$X$ theory can be obtained by the T-duality

\begin{equation}
g_{ij}(X)=\left(g^{ij}(\tilde{X})\right)^{-1}=\left[\frac{\sinh^{2}\left(\frac{1}{2}\frac{\tilde{\lambda}}{4\pi}\tilde{\mathcal{R}}\right)}{\frac{1}{4}\left(\frac{\tilde{\lambda}}{4\pi}\right)^{2}\mathcal{\tilde{\mathcal{R}}}^{2}}\right]^{a}\,_{i}\,\eta_{aj}.\label{eq:strong X}
\end{equation}

Moreover, our derivation also gives a simple way to manage the strong
$g_{s}$ coupling theories. With the identifications (\ref{eq:para iden}),
the $g_{s}<1$ theory is represented by (\ref{eq:Genus expansion})
or (\ref{eq:weak dual}), while the strong $g_{s}>1$ theory is described
by (\ref{eq:strong dual}), after substituting $\lambda=1/\tilde{\lambda}$
and $\mathbb{Y}=\mathbb{Y}(\tilde{\mathbb{Y}})$ from the T-duality
(\ref{eq:1st EOM}).

Or a even better choice is to consider the $X$ theory. Still with
the identifications (\ref{eq:para iden}), from the analysis above,
the $g_{s}<1$ theory must be (\ref{eq:strong X}), though it looks
quite different from the weak $g_{s}$<1 expression (\ref{eq:Genus expansion}) since it is expressed by $\tilde{\lambda}=1/\lambda$ and $\tilde{\mathbb{Y}}$.
The good news is that the strong $g_{s}>1$ theory is just (\ref{eq:weak X}),
already expressed in its native arguments. One can immediately conclude
that it is controlled by $1/\sqrt{g_{s}}$ \footnote{To get $1/\sqrt{g_{s}}$, we used an identification ${\mathbb{Y}}\sim {\cal W}$. It is possible to have other negative powers of $g_s$ depending on the specific identifications. But the power of $g_s$ should be negative.}. Therefore, as one may already
conceive, we really should expand the $g_{s}>1$ theory by $1/\sqrt{g_{s}}$,
in contrast to expanding $g_{s}<1$ theory by $g_{s}$.

\section{Some discussions}

There are several remarks we want to address

$\vphantom{}$

\noindent \underline{Strong genus expansion and supersymmetry} 

From our derivations, the perturbative genus expansion (\ref{eq:Genus expansion})
for $g_{s}\ll1$ is in agreement with the dual $\alpha'$ expansion
(\ref{eq:weak dual}) for $\sqrt{\alpha'}/R_{AdS}\gg1$. It is wonderful
that this is in perfect agreement with the prediction of AdS/CFT.
Moreover, we can predict the closed form of the strong genus expansion
for $g_{s}\gg1$! We can see in order to have the correspondence between
the two distinct expansions, supersymmetry is a must in the genus
expansion (\ref{eq:Genus expansion}). The GV in non-supersymmetric
background \cite{Gopakumar:1998ii,Gopakumar:1998jq} takes the form

\begin{equation}
\mathcal{F}=\int_{\epsilon}^{\infty}\frac{ds}{s}\mathrm{Tr}e^{-s\left(\triangle+m^{2}\right)}=\frac{1}{4}\int_{\epsilon}^{\infty}\frac{ds}{s}\frac{1}{\sinh\frac{seF}{2}}e^{-sm^{2}},
\end{equation}
which still maintains some similarity with the $\alpha'$ expansion
but we certainly need extra labor to arrive the non-supersymmetric
gauge theory. Another interesting problem is to extend the $\alpha'$
expansion to supersymmetric case and more accurate field identifications
may be derived.

$\vphantom{}$

\noindent \underline{Other geometries}

With the same pattern, it is straightforward to get the closed form
of local $\alpha'$ expansion for other non-trivial maximally symmetric
spaces. For De-Sitter space, one only needs to replace the sinh functions
by sine functions,

\begin{equation}
S_{dS}=-\frac{1}{4\pi}\int_{\Sigma}\partial\mathbb{Y}^{i}\partial\mathbb{Y}^{j}\left[\frac{\sin^{2}\left(\frac{1}{2}\frac{\lambda}{4\pi}\mathcal{R}\right)}{\frac{1}{4}\left(\frac{\lambda}{4\pi}\right)^{2}\mathcal{R}^{2}}\right]^{a}\,_{i}\,\eta_{aj}\,,\qquad\lambda\equiv\left(\frac{\sqrt{\alpha'}}{R_{dS}}\right)^{4}.
\end{equation}

\noindent For sphere $S_{n}$ and hyperbola $H_{n}$, we simply replace
the minkovski metric $\eta_{ij}$ by the Euclidean metric $\delta_{ij}$.
However, since there is no such a nice T-duality as (\ref{eq:1st EOM}),
it is tricky to talk about the dual theory. This might be the reason
why we have AdS/CFT only but not others. Nevertheless, we do have
a closed expression for the $\sqrt{\alpha'}/R_{c}$ expansion, it
is then natural to conjecture that for other maximally symmetric spaces
with $R_{c}\not=0$, the strongly coupled ($\sqrt{\alpha'}/R_{c}\gg1$)
worldsheet CFT still takes the form

\begin{equation}
S_{dS}=-\frac{1}{4\pi\alpha'}\int_{\Sigma}\partial X_{i}\partial X_{j}\left[\frac{\sin^{2}\left(\frac{1}{2}\frac{\tilde{\lambda}}{4\pi}\tilde{\mathcal{R}}\right)}{\frac{1}{4}\left(\frac{\tilde{\lambda}}{4\pi}\right)^{2}\mathcal{\tilde{\mathcal{R}}}^{2}}\right]^{a}\,_{i}\,\eta_{aj}\,,\qquad\tilde{\lambda}\equiv\left(\frac{R_{c}}{\sqrt{\alpha'}}\right)^{4}=\frac{1}{\lambda}.
\end{equation}

\noindent Note the $\alpha'$ in front of the integral, $X$ but not
$\tilde{X}$ outside the bracket and the barred quantities in the
bracket. Though it may not be possible to relate the parameters and
fields in two regimes, this closed expression for strongly coupled
system may have some applications. Moreover, even for asymmetric spaces,
we may be able to solve the strongly coupled systems by perturbation
on the corresponding maximally symmetric spaces.

%$\vphantom{}$
%
%
%\noindent \underline{Why locally closed expression?}
%
%It may look confusing why it is necessary to have such a closed form
%defined locally only. Anyway, we already have a globally defined closed
%form for the metric $g_{\mu\nu}=\left(\frac{R_{AdS}}{Z}\right)^{2}\eta_{\mu\nu}$.
%An immediate reason is that the global metric
%$g_{\mu\nu}=\left(\frac{R_{AdS}}{Z}\right)^{2}\eta_{\mu\nu}$ does
%not manifest the explicit $\alpha'$ dependence. We are thus unable
%to analyze the $\alpha'$ expansion. 
%
%This puzzle can be rephrased in another way: How is possible a local geometry (local $\alpha'$ expansion)  so closely related to topologies (genus expansion)? A rough interpretation is that, bearing in mind that both AdS and flat spacetime are maximally symmetric space (MSS), local geometry already indicates some global properties. The specific $\alpha'$ expansion depends on the coordinates but the sum, i.e. the metric as a geometric object, is independent of the coordinate system. The mathematics behind our results is that local geometry of MSS 
%
%On the other hand,  ``Gauge'' a theory usually means localize the theory.
%As bundles defined locally on a manifold, the most natural and efficient
%way to study Yang-Mills theory is to use local frames. Follow this
%strategy, there are numerous efforts trying to unify Yang-Mills theory
%and gravitation on the local basis. Thus, one can anticipate that in order to gain connections
%with Yang-Mills theory, RNC can make things more explicit and simpler, as we showed.

$\vphantom{}$

\noindent \underline{Ooguri-Vafa formula and open string}

We addressed the GV formula and closed string in this paper. In Ooguri-Vafa
formula \cite{Ooguri:1999bv}, which is an open string extension of
GV, an extra D4-brane in $\mathbb{R}^{2}\times L$, $L\subset Y$
as a special Lagrangian manifold, is introduced and the supersymmetry
is broken to leave at most four supercharges. It would be of great
important to generalize our discussion in this paper to that situation
since one can expect more information about the non-abelian gauge
theory can be extracted.

$\vphantom{}$

\noindent \underline{Moduli calculation}

Determining the moduli space of higher genus surfaces is a very hard
problem and no systematical method is available. The equivalence between
the $\alpha'$ expansion and genus expansion provides a possible way
to attack this problem. It is quite interesting that the closed expression
of the $\alpha'$ expansion is obtained locally, while the genus expansion
is not restricted to local region. It would be of interest to test
it by calculating $n$-point functions on non-trivial surfaces, say,
annulus or torus. 

$\vphantom{}$

\noindent \underline{Hodge integrals and topological string theory}

It is obvious to see that our results are related to the generating
function for Hodge integrals:

\begin{equation}
\mathcal{F}\left(t,k\right)\equiv1+\underset{g\geq1}{\sum}t^{2g}\underset{i=0}{\overset{g}{\sum}}k^{i}\int_{\bar{M}_{g,1}}\psi^{2g-2+i}\lambda_{g-i}=\left(\frac{t/2}{\sin\left(t/2\right)}\right)^{k+1},
\end{equation}

\noindent where $t$ and $k$ are some parameters, $\bar{M}_{g,n}$
is the moduli space with genus $g$ and $n$ distinct marked points,
$\psi^{i}$ is the first Chern class for a marked $i$ cotangent line
bundle, and $\lambda_{j}$ is $j$th Chern class of Hodge bundle.
This Hodge integrals play a central role in the Gromov-Witten theory
and topological string theory. It is readily to see that $F\left(t,1\right)$
is T-dual to $F\left(t,-3\right)$ on the basis of our results.

\vspace{5mm}

\noindent {\bf Acknowledgements} 
%We would like to acknowledge illuminating discussions with. 
This work is supported in part by the NSFC (Grant No. 11175039 and 11375121).  We would like to acknowledge helpful discussions with B. Feng, Y. He, S. Kim, H. Nakajima and D. Polyakov. We are indebted to S.-T. Yau for useful instructions. H. Wu is indebted to D. Jafferis, D. Xie and W. Yan for comments on the draft.
%H. Y. is grateful to the hospitality of the Institute of Theoretical Physics, Chinese %Academy of Sciences where part of this work is done.

\appendix

\section{Appendix}

\begin{center}
A Closed Formula of Metric Expansion for Maximally Symmetric Spaces in RNC 
\end{center}

In this appendix, we will calculate the metric expansion in Riemann normal
coordinates (RNC). This normal coordinate system on a Riemannian manifold
centered at $\bar{x}$ is defined by%
\begin{equation}
g_{ij}\left(  \bar{x}\right)  =\eta_{ij}\text{, }\Gamma_{ij}^{k}\left(
\bar{x}\right)  =0.
\end{equation}
Consider a point $X$ in a neighborhood of $\bar{x}$, whose coordinates are
$\bar{x}+y$ in the RNC. The geodesic connecting $\bar{x}$ with $X$ is then
given by $\gamma^{i}\left(  s\right)  =\bar{x}^{i}+y^{i}s$ with $\gamma\left(
0\right)  =\bar{x}$ and $\gamma\left(  1\right)  =X$. By substituting it into
the geodesic equation, this implies that%
\begin{equation}
\Gamma_{ij}^{k}\left(  X\right)  y^{i}y^{j}=0\text{.}\label{eq:Gammayy}%
\end{equation}
Taylor expanding LHS of eqn. $\left(  \ref{eq:Gammayy}\right)  $ in terms of
$y$ at $\bar{x}$ leads to%
\begin{equation}
\Gamma_{(ij,i_{1}\cdots i_{n})}^{k}\left(  \bar{x}\right)  =0\text{, for
}n\geq0\text{.}\label{eq:GammaI}%
\end{equation}

Since the metric $g_{ij}$ can be expressed with respect to the vielbein
$e_{i}^{a}$:%
\[
g_{ij}\left(  X\right)  =e_{i}^{a}\left(  X\right)  e_{j}^{b}\left(  X\right)
\eta_{ab},
\]
we could first find an expression for $e_{i}^{a}\left(  X\right)  $ in terms
of the curvature. To do so, we can choose synchronous gauge for the vielbein:%
\begin{equation}
i_{\mathbf{y}}e_{i}^{a}\left(  X\right)  =\delta_{i}^{a}y^{i}%
,\label{eq:vielGC}%
\end{equation}
where the radial vector $\mathbf{y}$ is $\mathbf{y}=y^{i}\partial_{i}$, and
$i_{\mathbf{y}}$ is the interior product. In RNC, the gauge condition $\left(
\ref{eq:vielGC}\right)  $ implies a condition for the vielbein connection:%
\begin{equation}
i_{\mathbf{y}}\omega^{ab}\left(  X\right)  =0.
\end{equation}
In fact, differentiating both sides of eqn. $\left(  \ref{eq:vielGC}\right)  $
gives%
\[
y^{i}y^{j}\partial_{i}e_{j}^{a}\left(  X\right)  =0,
\]
where we use $e_{j}^{a}\left(  0\right)  =\delta_{j}^{a}$. Since the
connection $\Gamma_{jk}^{i}$ is
\begin{equation}
\Gamma_{jk}^{i}\left(  X\right)  =e_{a}^{i}\left(  X\right)  e_{k,j}%
^{a}\left(  X\right)  +e_{a}^{i}\left(  X\right)  \omega_{j\text{ }d}^{\text{
}a}\left(  X\right)  e_{k}^{d}\left(  X\right)  ,
\end{equation}
eqn. $\left(  \ref{eq:Gammayy}\right)  $ leads to%
\begin{equation}
\delta_{j}^{b}y^{j}\left(  y^{i}\omega_{i\text{ }b}^{\text{ }a}\right)
=0\Rightarrow i_{\mathbf{y}}\omega^{ab}=y^{i}\omega_{i}^{\text{ }ab}=0\text{.}%
\end{equation}
By Taylor expanding $i_{\mathbf{y}}\omega^{ab}\left(  X\right)  =0$ at
$\bar{x}$, we find
\begin{equation}
\omega_{(i,i_{1}\cdots i_{n})}^{\text{ }ab}\left(  0\right)  =0\text{.}%
\label{eq:OmegaI}%
\end{equation}

Using $\mathcal{L}_{\mathbf{y}}=i_{\mathbf{y}}d+di_{\mathbf{y}}$, one finds%
\begin{align}
\mathcal{L}_{\mathbf{y}}e^{a}  &  =d\left(  i_{\mathbf{y}}e^{a}\right)
+\left(  i_{\mathbf{y}}e^{b}\right)  \omega_{\text{ }b}^{a},\label{eq:Le}\\
\mathcal{L}_{\mathbf{y}}\mathcal{L}_{\mathbf{y}}e^{a}  &  =\left(
i_{\mathbf{y}}e^{b}\right)  \mathcal{L}_{\mathbf{y}}\omega_{\text{ }b}%
^{a}+\omega_{\text{ }b}^{a}\mathcal{L}_{\mathbf{y}}i_{\mathbf{y}}%
e^{b}+\mathcal{L}_{\mathbf{y}}di_{\mathbf{y}}e^{a}, \label{eq:LLe}%
\end{align}
where we use $de^{a}+\omega_{\text{ }b}^{a}\wedge e^{b}=0$ since the torsion
is absent. The gauge condition $\left(  \ref{eq:vielGC}\right)  $ gives%
\begin{align}
\mathcal{L}_{\mathbf{y}}di_{\mathbf{y}}e^{a}  &  =d\left(  i_{\mathbf{y}}%
e^{a}\right)  \text{,}\nonumber\\
\mathcal{L}_{\mathbf{y}}\left(  i_{\mathbf{y}}e^{b}\right)   &  =i_{\mathbf{y}%
}e^{b}\text{,}%
\end{align}
and hence subtracting eqn. $\left(  \ref{eq:LLe}\right)  $ from eqn. $\left(
\ref{eq:Le}\right)  $ leads to%
\[
\left(  \mathcal{L}_{\mathbf{y}}-1\right)  \mathcal{L}_{\mathbf{y}}%
e^{a}=\left(  i_{\mathbf{y}}e^{b}\right)  \mathcal{L}_{\mathbf{y}}%
\omega_{\text{ }b}^{a}.
\]
Then using the gauge condition $i_{\mathbf{y}}\omega^{ab}=0$ and
\begin{equation}
R_{\text{ }b}^{a}=\frac{1}{2}R_{\text{ }bij}^{a}dx^{i}\wedge dx^{j}%
=d\omega_{\text{ }b}^{a}+\omega_{\text{ }c}^{a}\wedge\omega_{\text{ }b}^{c},
\end{equation}
we have%
\begin{equation}
\left(  y\cdot\partial\right)  \left(  y\cdot\partial+1\right)  e_{i}%
^{a}\left(  y\right)  =y^{j}y^{k}R_{\text{ }jkb}^{a}\left(  y\right)
e_{i}^{b}\left(  y\right)  \text{,} \label{eq:dy}%
\end{equation}
where $e_{i}^{a}\left(  y\right)  =e_{i}^{a}\left(  X\right)  $. Noting that
\begin{equation}
\frac{d^{n}}{dv^{n}}e_{i}^{a}\left(  vy\right)  =y^{i_{1}}\cdots y^{i_{n}%
}\partial_{i_{1}}\cdots\partial_{i_{n}}e_{i}^{a}\left(  vy\right)  ,
\end{equation}
we can rewrite eqn. $\left(  \ref{eq:dy}\right)  $ as a differential equation
in terms of $v$%
\begin{equation}
\frac{d}{dv}\left[  v^{2}\frac{d}{dv}e_{i}^{a}\left(  vy\right)  \right]
=v^{2}y^{j}y^{k}R_{\text{ }jkb}^{a}\left(  vy\right)  e_{i}^{b}\left(
vy\right)  ,
\end{equation}
which can be integrated to%
\begin{equation}
e_{i}^{a}\left(  y\right)  =\delta_{i}^{a}+\int_{0}^{1}\left(  1-v\right)
vy^{j}y^{k}R_{\text{ }jkb}^{a}\left(  vy\right)  e_{i}^{b}\left(  vy\right)
dv\text{.} \label{eq:eSol}%
\end{equation}
It is noteworthy that the solution $\left(  \ref{eq:eSol}\right)  $ indeed
satisfies the gauge condition $\left(  \ref{eq:vielGC}\right)  $ since
$R_{\text{ }jki}^{a}y^{k}y^{i}=0$.

Introducing the abbreviation%
\begin{equation}
R_{\text{ }b}^{a}\left(  z,y\right)  =R_{\text{ }ijb}^{a}\left(  z\right)
y^{i}y^{j},
\end{equation}
we can express $e_{i}^{a}$ in terms of $R_{\text{ }b}^{a}$ by iterating eqn.
$\left(  \ref{eq:eSol}\right)  $%
\begin{align}
e_{i}^{a}\left(  y\right)   &  =\delta_{i}^{a}+%
%TCIMACRO{\dsum \limits_{n=1}}%
%BeginExpansion
{\displaystyle\sum\limits_{n=1}}
%EndExpansion
\int_{0}^{1}dv_{1}\left(  1-v_{1}\right)  \int_{0}^{1}dv_{2}\left(
1-v_{2}\right)  \cdots\int_{0}^{1}dv_{n}\left(  1-v_{n}\right)  \nonumber\\
&  \left(  v_{1}^{2n-1}\cdots v_{n-1}^{3}v_{n}\right)  R_{\text{ }l_{1}}%
^{a}\left(  v_{1}y,y\right)  R_{\text{ \ }l_{2}}^{l_{1}}\left(  v_{1}%
v_{2}y,y\right)  \cdots R_{\text{ \ \ \ \ \ }b}^{l_{n-1}}\left(  v_{1}\cdots
v_{n}y,y\right)  \delta_{i}^{b}.
\end{align}
On the other hand, Taylor expanding the curvature tensor in $R_{\text{ }b}%
^{a}\left(  ty,y\right)  $ gives%
\begin{align}
R_{\text{ }b}^{a}\left(  ty,y\right)   &  =%
%TCIMACRO{\dsum \limits_{p=0}^{\infty}}%
%BeginExpansion
{\displaystyle\sum\limits_{p=0}^{\infty}}
%EndExpansion
\frac{t^{p}\left(  y^{i}\partial_{z^{i}}\right)  ^{p}}{p!}R_{\text{ }b}%
^{a}\left(  z,y\right)  |_{z=0},\nonumber\\
&  =%
%TCIMACRO{\dsum \limits_{p=0}^{\infty}}%
%BeginExpansion
{\displaystyle\sum\limits_{p=0}^{\infty}}
%EndExpansion
\frac{t^{p}\left(  y^{i}\nabla_{z^{i}}\right)  ^{p}}{p!}R_{\text{ }b}%
^{a}\left(  z,y\right)  |_{z=0},\label{eq:Rp}%
\end{align}
where we use eqns. $\left(  \ref{eq:GammaI}\right)  $ and $\left(
\ref{eq:OmegaI}\right)  $ to replace $\partial$ with $\nabla$ in the second line.

We now focus on maximally symmetric spaces, where $\nabla_{k}R_{\text{ }%
ijl}^{m}=0$. Using eqn. $\left(  \ref{eq:Rp}\right)  $, we find that
\begin{equation}
R_{\text{ }b}^{a}\left(  ty,y\right)  =R_{\text{ }b}^{a}\left(  0,y\right)
\end{equation}
in maximally symmetric spaces, and hence%
\begin{equation}
e_{i}^{a}\left(  y\right)  =%
%TCIMACRO{\dsum \limits_{n=0}}%
%BeginExpansion
{\displaystyle\sum\limits_{n=0}}
%EndExpansion
\frac{\left[  R^{n}\left(  0,y\right)  \right]  _{\text{ }b}^{a}\delta_{i}%
^{b}}{\left(  2n+1\right)  !}\text{.}%
\end{equation}
Then, the result for the metric $g_{ij}=e_{i}^{a}e_{j}^{b}\delta_{ab}$
becomes
\begin{equation}
g_{ij}\left(  X\right)  =\eta_{ij}+\frac{1}{2}%
%TCIMACRO{\dsum \limits_{n=1}}%
%BeginExpansion
{\displaystyle\sum\limits_{n=1}}
%EndExpansion
\frac{2^{2n+2}\eta_{im}R\left(  0,y\right)  _{\text{ \thinspace\ }a_{1}}%
^{m}R\left(  0,y\right)  _{\text{ \thinspace\ \ }a_{2}}^{a_{1}}\cdots R\left(
0,y\right)  _{\text{ \thinspace\ \ \ \ \ \ }j}^{a_{n-1}}}{\left(  2n+2\right)
!}.\label{eq:gs}%
\end{equation}
For maximally symmetric spaces, the Riemann curvature tensor is
\begin{equation}
R_{ikjl}=L^{-2}\left(  g_{ij}g_{kl}-g_{il}g_{kj}\right)  .
\end{equation}
If $X=\bar{x}+\ell_{s}\mathbb{Y}$ where $\mathbb{Y}$'s are dimensionless, we
find%
\begin{equation}
R_{\text{ }b}^{a}\left(  0,\ell_{s}\mathbb{Y}\right)  =\left(  -1\right)
^{\frac{\text{sgn}\left(  L^{2}\right)  +1}{2}}\frac{1}{4}\left(
\frac{\lambda}{4\pi}\right)  ^{2}\left(  \mathcal{R}^{2}\right)  _{\text{ }%
b}^{a},
\end{equation}
where we define%
\begin{equation}
\lambda\equiv\frac{\ell_{s}^{4}}{L^{4}}\text{, }\left(  \mathcal{R}%
^{2}\right)  _{\text{ }b}^{a}\equiv64\pi^{2}\left(  \frac{\left\vert
L\right\vert }{\ell_{s}}\right)  ^{6}\left(  \delta_{b}^{a}\mathbb{Y}%
^{2}-\mathbb{Y}^{a}\mathbb{Y}_{b}\right)  \text{.}%
\end{equation}
Summing up the series $\left(  \ref{eq:gs}\right)  $, we find%
\begin{equation}
g_{ij}\left(  X\right)  =\left[  \frac{\sinh^{2}\left(  \left(  -i\right)
^{\frac{\text{sgn}\left(  L^{2}\right)  +1}{2}}\frac{1}{2}\frac{\lambda}{4\pi
}\mathcal{R}\right)  }{\left(  -1\right)  ^{\frac{\text{sgn}\left(
L^{2}\right)  +1}{2}}\frac{1}{4}\left(  \frac{\lambda}{4\pi}\right)
^{2}\mathcal{R}^{2}}\right]  _{i}^{a}\eta_{aj}.\label{eq:gx}%
\end{equation}
For example, one has $L^{2}=-R_{AdS}^{2}$ in a D-dimensional AdS spacetime,
and eqn. $\left(  \ref{eq:gx}\right)  $ becomes eqn. $\left(  \ref{eq:weak X}%
\right)  $.

\end{document}